# Topology optimized multi-functional mechanically reconfigurable meta-optics studied at microwave frequencies

CONNER BALLEW, GREGORY ROBERTS, PHILIP CAMAYD-MUÑOZ, MAXIMILIEN F DEBBAS, AND ANDREI FARAON[*]

[1]*Kavli Nanoscience Institute and Thomas J. Watson Sr. Laboratory of Applied Physics, California Institute of Technology, Pasadena, California 91125, USA*
*\*faraon@caltech.edu*

**Abstract:** Metasurfaces advanced the field of optics by reducing the thickness of optical components and merging multiple functionalities into a single layer device. However, this generally comes with a reduction in performance, especially for multifunctional and broadband applications. Three-dimensional metastructures can provide the necessary degrees of freedom for advanced applications, while maintaining minimal thickness. This work explores 3D mechanically reconfigurable devices that perform focusing, spectral demultiplexing, and polarization sorting based on mechanical configuration. As proof of concept, a rotatable device, auxetic device, and a shearing-based device are designed with adjoint-based topology optimization, 3D-printed, and measured at microwave frequencies (7.6-11.6 GHz) in an anechoic chamber.

## 1. Introduction

Traditional optical design combines independent bulk elements to achieve complex functionality. Recent advances in nano-fabrication technologies enabled the miniaturization of bulk components by synthesizing multiple optical functionalities into single, subwavelength-thick layers called metasurfaces [1,2]. However, the performance of metasurfaces is limited by the number of optical modes that can be controlled, which scales with the volume of the device and the maximum refractive index contrast [3,4].

Systems of metasurfaces have been used to perform functions that are beyond the capabilities of a single metasurface. Some novel platforms for this technique include fabricating two metasurfaces on opposite sides of a transparent wafer [5,6]. A "folded-optics" platform uses reflective metasurfaces on both sides of a wafer to enable many interactions with different metasurfaces in a small, wafer-thick region [7]. This technique still uses a modular design approach, requiring the different metasurfaces be spatially separated to mitigate the unaccounted-for effects of multiple scattering events between elements that effectively invalidate the local phase approximation that has made metasurface design tractable [8].

To expand the functionality of metaoptics, adjoint-based topology optimization is capable of designing high-performing dielectric structures with nonintuitive index distributions and objective functions that are challenging to achieve with traditional methods [9,10]. While much of the work has been on 2D platforms such as silicon photonic waveguides [11], free-space 2.5D and 3D devices have also been explored recently [12–16]. In a previous work, a passive 3D device was designed that functions as an red-green-blue (RGB) color splitter and polarizer, scaled up to microwave frequencies [13]. An interesting prospect for this type of device is to expand its functionality while minimizing any corresponding reduction in efficiency. For a passive device, multiple dependent functionalities will necessarily compete for efficiency. For example, the efficiency of demultiplexing spectral bins will be reduced if the device simultaneously sorts polarization. One way to maintain high efficiencies amidst multiple competing functionalities is to design active device that can switch functionalities in time to perform the different tasks.

In this work, we use inverse design techniques and topology optimization to design devices capable of switching between optical functions through mechanical reconfiguration. We explore 3 different methods of reconfigurability: a rotatable device that switches between three-band spectral splitting and broadband focusing when rotated 180°; a negative Poisson ratio, or "auxetic", device based on rotating rigid squares that performs the same functions as the rotatable device; and a device based on shearing adjacent layers that can switch between three-band spectral splitting, broadband focusing, and polarization sorting.

## 2. Measurement setup and design methods

### 2.1 Measurement setup

Although 3D devices can be fabricated at the micro-scale with methods such as two-photon direct laser writing and greyscale lithography, it remains challenging to fabricate fully 3D structures with the required features sizes and alignment accuracy for optical and near-infrared applications [17]. For this work, we test our ideas in the microwave range from 7.6 to 11.6 GHz where Fused Filament Fabrication 3D-printing with polylactide acid (PLA) can generate subwavelength feature sizes with minimal material absorption. The 42% fractional bandwidth of this frequency range is the same fractional bandwidth as the range of 450 nm to 690 nm light, which is nearly the full visible spectrum.

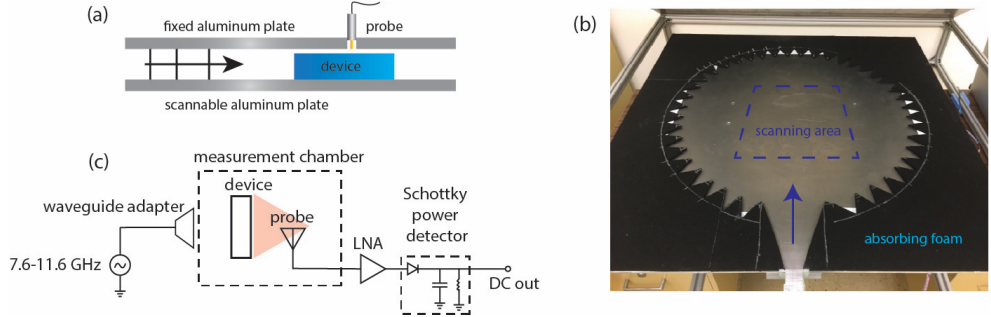

Fig. 1. The anechoic chamber used to measure the 3D printed devices. (a) A sideview of the chamber. A TEM mode is injected into the anechoic chamber, exciting the device at approximately normal incidence. The fields in the cavity, including fields above the device, are measured with a small antenna probe extending through the top aluminum plate. The bottom aluminum plate is scanned in two directions to make a full 2D scan. (b) A picture of the anechoic chamber with the top aluminum plate removed. The scanning area is surrounded by absorbing foam. Triangles were cut out of the absorbing foam to reduce reflections. (c) A schematic representation of the measurement setup. The microwave source (Windfreak SynthHD) is scanned from 7.6 GHz to 11.6 GHz and is injected through a waveguide adapter. The signal from the measurement probe is first amplified with two low-noise amplifiers, and the signal is then passed through a Schottky power detector (Keysight 8470B). The output is proportional to the field amplitude in the cavity.

We use a homemade anechoic microwave chamber to measure the electric fields of the devices, shown in Fig. 1. The design and construction of this setup was inspired by the work in reference [18]. The measurement system consists of a parallel plate waveguide supporting only the fundamental TEM mode in the frequency range of interest when propagating through air or PLA. A sideview schematic of the system is shown in Fig. 1(a), and the constructed system with the top plate removed is shown in Fig. 1(b). A circuit view of the system is shown in Fig. 1(c). The output of a microwave synthesizer (Windfreak SynthHD) is injected into the chamber through a waveguide feed. This source is scanned from 7.6 GHz to 11.6 GHz, which is the designed bandwidth of operation for all presented devices. A small probe antenna (SMA Connector Receptacle) is extended through the top plate of the chamber to probe the fields inside the cavity. The probe antenna signal is amplified with low noise amplifiers and propagated through a Schottky diode detector (Keysight 8470B). The resulting signal is proportional to the amplitude of the fields in the cavity. The bottom plate is moved with stepper motors to make a 2D scan of the field amplitude inside the chamber. The probe antenna does not extend into the cavity, so the field intensity in the small air gap above the device can be measured. These fields are approximately proportional to the fields within the device material [18].

An important characteristic of this test setup is it measures 2D devices under TE-polarized illumination. The response of the device in the microwave chamber is nearly identical to the response of a device placed in a free space with the same index distribution, assuming the index distribution is extruded infinitely in the direction separating the waveguide plates [18]. This simplification from 3D to 2D implies we can optimize the device in a 2D coordinate system, thus drastically reducing the overall time of the optimization. Although these devices can be fully described by a 2D coordinate system, they are analogous to 3D fabricated devices since patterning occurs in the direction of light propagation.

We design our devices assuming a refractive index of 1.5. While the actual refractive index of PLA is closer to 1.65 [19], the devices only fill about 80% of the gap in the microwave test chamber. The effective index is thus approximately 1.5. There is uncertainty in this value which likely contributes to discrepancies between simulated and measured device performance. One source of the uncertainty is predicted to originate from the layered nature of the 3D print which has previously been reported to cause a 7% anisotropy in PLA permittivity, with the permittivity ranging from 2.75 in the direction of normal to the printing surface and 2.96 in the direction parallel to the printing surface [19]. Furthermore, if the permittivity depends on the layered nature of the print, then it is likely affected by the 3D printer quality and layer height, which may be different from the references used to estimate the PLA permittivity.

## 2.2 Topology optimization

The three designs presented are all optimized with the same strategy. The optimization technique is similar to the techniques presented in [13,20], with some differences to account for the mechanical reconfiguration of each device, which will be briefly summarized here. The adjoint variable method is a technique to efficiently compute the gradient of a figure of merit with respect to a design region permittivity by combining the results of just two simulations, a "forward" simulation and a backward, or "adjoint", simulation. We use a time-domain Maxwell's equations solver (Lumerical FDTD), so each simulation contains results for the entire bandwidth of interest.

The optimization begins with each device modelled as a block of material with a permittivity between that of air and PLA. The gradient of every figure of merit is computed for each of its potential configurations. All gradients are combined with a weighted average, with weights chosen according to Eq. 1 such that all figure of merits seek the same efficiency. In this equation, *FoM* represents the current value of a figure of merit, *N* is the total number of figures of merit, and $w_i$ represents the weigh applied to its respective merit function's gradient. The maximum operator is used to ensure the weights are never negative, thus ignoring the gradient of high-performing figures of merit rather than forcing the figure of merit to decrease. The 2/N factor is used to ensure all weights conveniently sum to 1, unless some weights were negative before the maximum operation.

(1) $$w_i = \max\left(\frac{2}{N} - \frac{FoM_i^2}{\Sigma FoM_j^2}, 0\right)$$

The permittivity is stepped in the direction of this weighted gradient, and the process is repeated until all pixels of the design region have become either PLA or air. It is not guaranteed that all pixels will ultimately become air or PLA, since they can settle at a local extreme point at a fictitious permittivity in between PLA and air. Thus, we conclude the optimization by forcing each pixel to be either PLA or air which has a negligible effect on the performance of the device. Once the design is finalized, the resulting permittivity distribution is exported to an .STL file and 3D printed in PLA using an Ultimaker S3.

## 3. Results

Here we describe the results from the three different devices. Measurement results feature a 2D scan of the microwave cavity chamber for each configuration, with the red-green-blue colors representing the equivalent hue of visible light when the microwave frequencies are scaled by a factor of ×59,618. The microwave components have varying scattering parameters over the bandwidth of interest, a scan of the empty cavity is used to normalize device measurements to the total power injected into the cavity. Each device focuses to one or more of three separate pixels, depending on the desired functionality of the configuration. The pixels are depicted as either red, green, or blue, indicating the assigned color in the spectral splitting configuration with red as the lowest frequency bin (7.6-8.9 GHz), green as the middle frequency bin (8.9-10.2 GHz), and blue as the high frequency bin (10.2-11.6 GHz). These frequency bins are referred to simply as red, green, or blue from now on.

The focal plane is analyzed in simulation to determine the sorting efficiency, defined as the fraction of incident power transmitting through the target pixel. To quantitatively corroborate the measured intensity profile with the simulated intensity profile, the measured devices feature a series of plots that compare the normalized intensity integrated over the frequency bands of interest for each functionality.

### 3.1 Rotatable device

The first device presented here changes its function by undergoing a 180° rotation. The device is designed only for TE light. It focuses broadband light to the center pixel when illuminated by a normally-incident planewave from one side, and focuses red, green, and blue light to the respectively colored pixels when illuminated from the other side, as shown in Fig. 2(a-d). The footprint of the device is 6.2 cm × 18.6 cm, which is $2\lambda \times 6\lambda$ at the center wavelength of 3.1 cm, shown in Fig. 2(e). To ensure structural stability the device has a thin frame of PLA, and a connectivity constraint is enforced every ten iterations of the optimization so that the index distribution converges to a fully connected structure.

This device is uniquely simple among the devices presented, since a 180° rotation does not alter the scattering matrix beyond a transposition due to reciprocity. A reasonable concern is that this device will be limited in performance because of this. However, reciprocity does not strongly affect this device, since inputs from both sides are assumed to be normally incident. Thus, only scattering components mapping a normally incident input to a normally incident output are coupled by reciprocity, while the desired output fields are comprised of many more uncoupled planewave components.

The sorting efficiency as a function of frequency is shown in Fig. 2(f). The sorting efficiency averaged across the relevant frequencies for each function is 50.6%. This outperforms a traditional three-pixel absorptive Bayer filter arrays, which have a maximum theoretical sorting efficiency of 33% for each frequency band. To quantitatively compare the intensity profiles of measured and simulated results, normalized intensity profiles are shown in Fig. 2(g) for each function.

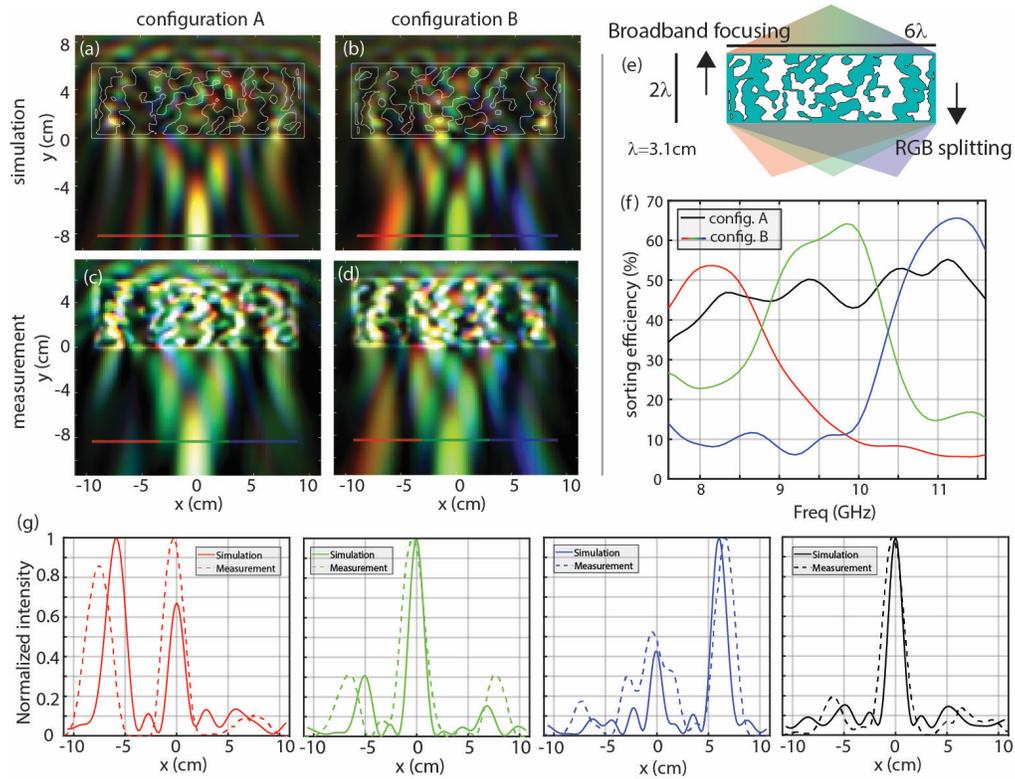

Fig. 2. Rotatable device performing broadband focusing in one configuration and spectral splitting in the other configuration. (a) Broadband focusing simulation, (b) spectral splitting simulation, (c) broadband focusing measurement, and (d) spectral splitting measurement. (e) Schematic of the device, where blue represents PLA and white represents air. The footprint of the device is 6.2 cm × 18.6 cm. (f) Analysis of the simulated fields at the focal plane showing the sorting efficiency for each function, defined as the fraction of incident power reaching the target pixel. Red, green, and blue light are focused to their respectively colored pixels, and sorting efficiency is drawn in red, green, and blue, respectively. The broadband focusing function focuses all light to the middle pixel and is drawn in black. (g) Comparison of simulated and measurement normalized intensity profiles at the focal plane. Configuration B intensities are integrated over (left) 7.6 to 8.9 GHz, (center-left) 8.9 GHz to 10.2 GHz, (center-right) 10.2 to 11.6 GHz. (right) Configuration A intensity integrated from 7.6 to 11.6 GHz.

### *3.2 Auxetic device*

Auxetic metameterials have a negative Poisson's ratio – they have the nonintuitive property of widening in the transverse direction when stretched and narrowing in the transverse direction when compressed. Such metamaterials are useful for tailoring the mechanical properties of devices, and can increase indentation resistance, shear resistance, energy absorption, hardness, and fracture toughness [21]. They can be fabricated at the macro-scale through techniques such as 3D printing and have been studied at the micro- and nanoscale [22,23]. Here we present a device capable of switching its optical functionality through the well-studied auxetic transformation of rotating rigid squares, yielding devices with a Poisson's ratio of -1 [24].

The two functions chosen here are broadband focusing and spectral splitting for TE polarization – the same as the rotatable device. The first configuration features a 0° rotation of all squares, while the second configuration features a ±90° of each square, with each square having an opposite angular rotation as its neighboring squares. The device is fully connected, with a frame around each square element to ensure structural stability. The device and its auxetic transformation are shown in Fig. 3(e).

A total of three devices were fabricated. The first device demonstrates the auxetic transformation. To fabricate this, the 3D-print was paused at half the thickness of the device, a nylon mesh was manually inserted, and the 3D-print was then resumed to completion. After the print, the nylon mesh was cut to leave only the required flexible hinges between each square. A video of this device being manually actuated is available in Visualization 1. The other two devices are the two different configurations printed directly, which were then measured.

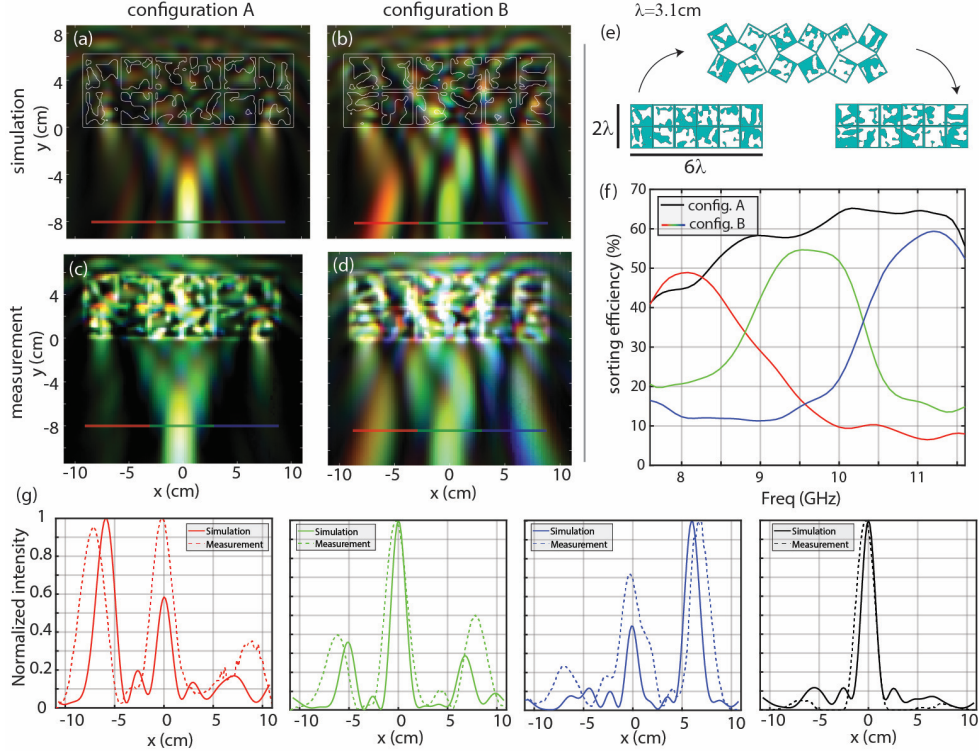

Fig. 3. Auxetic device performing broadband focusing in one configuration and spectral splitting in the other configuration. (a) Broadband focusing simulation, (b) spectral splitting simulation, (c) broadband focusing measurement, and (d) spectral splitting measurement. (e) Demonstration of the auxetic transformation and device footprint, where blue represents PLA and white represents air. The footprint of the device is 6.2 cm × 18.6 cm. (f) Analysis of the focal plane of this device, showing the sorting efficiency for each function, defined as the fraction of incident power reaching the target pixel. Red, green, and blue light are focused to their respectively colored pixels, and sorting efficiency is drawn in red, green, and blue, respectively. The broadband focusing function focuses all light to the middle pixel and is drawn in black. (g) Comparison of simulated and measurement normalized intensity profiles at the focal plane. Configuration B intensities are integrated over (left) 7.6 to 8.9 GHz, (center-left) 8.9 GHz to 10.2 GHz, (center-right) 10.2 to 11.6 GHz. (right) Configuration A intensity integrated from 7.6 to 11.6 GHz.

The auxetic design and rotatable design have the same device size and focal length, so the performance of the two devices can be directly compared. The simulated sorting efficiency is shown in Fig. 3(f). Since the auxetic device has a frame around each square element, less of the design area is available for optimization. This detracts from the degrees of freedom of the device, which may explain why the device has a 48.0% average sorting efficiency, 2.6% less than the rotatable device. The difference in average efficiency may also arise from the different natures of the mechanical reconfiguration. The simulated and measured intensities within the full test chamber are shown in Fig 3(a-d). To quantitatively compare the intensity profiles of measured and simulated results, normalized intensity profiles are shown in Fig. 3(g) for each function.

*3.3 Shearing device*

The final device features a transformation based on shearing alternate layers. This action is achievable at the microscale through MEMS electrostatic actuation. Unlike the previous devices, this device switches between three different functionalities: spectral splitting shown in Fig. 4(a), broadband focusing shown in Fig. 4(b), and polarization splitting shown in Fig. 4(c). All functionalities are designed for both TE and TM polarizations.

The added polarization control demands more degrees of freedom than the rotatable and auxetic devices, and it was found that the device needed to be nearly 3× thicker than the rotatable and auxetic devices to achieve satisfactory performance of 59.2% average sorting efficiency. This device does not have a supporting frame and does not enforce connectivity like the previous devices.

This device was analyzed only in simulation due to limitations within the measurement system: the measurement chamber only supports a TE-polarized TEM mode, and the scannable region is too small to measure this device. A 4-layer device was designed and tested with similar agreement between simulation and measurement to the rotatable and auxetic devices, but the device could not achieve all objective functions. Data for this device is available on request.

The configurations and simulation results for the 8-layer simulated device are summarized in Fig. 4. The device is a stack of eight 8 cm × 2 cm layers. The mechanical actuation displaces adjacent layers in equal and opposite directions by 3 cm, which is approximately one wavelength. The first configuration, shown in Fig. 4(a), shows spectral splitting behavior with 42% average sorting efficiency for TE and 40% average sorting efficiency for TM. The crosstalk between the different spectral bins is worse than the rotatable and auxetic devices, with the worst case occurring for the TM-polarized green input which focuses only 1.2× more power to the desired green pixel than the undesired blue and red pixels.

The neutral position of the device, in which all layers are aligned as shown in Fig. 4(b), features broadband focusing with high efficiency. The aperture size of this configuration is smaller than the other two configurations, and the sorting efficiency is normalized to the power incident on this smaller aperture when analyzing this configuration. The sorting efficiency shown in Fig. 4(e) is expected to be uniformly high across all frequencies and both polarizations.

The final configuration sorts TE polarization to the leftmost pixel and TM polarization to the center pixel. The transmission through each pixel, averaged across the entire spectrum, is summarized in matrix form in Fig. 4(f): a TE input focuses 47% power to the correct pixel and 28% power to the incorrect pixel, the TM input focuses 62% power to the correct pixel and 24% power to the incorrect pixel. More power is coupled to the desired pixel than the undesired pixel at all frequencies except for the case of TE input frequencies below approximately 9 GHz. In this case more power is directed towards the incorrect green pixel than the correct red pixel. It is possible that this could be fixed by sacrificing performance in the other functionalities, such as by tuning the weighting scheme described in Eq. 1, or by increasing the thickness or index contrast of the device.

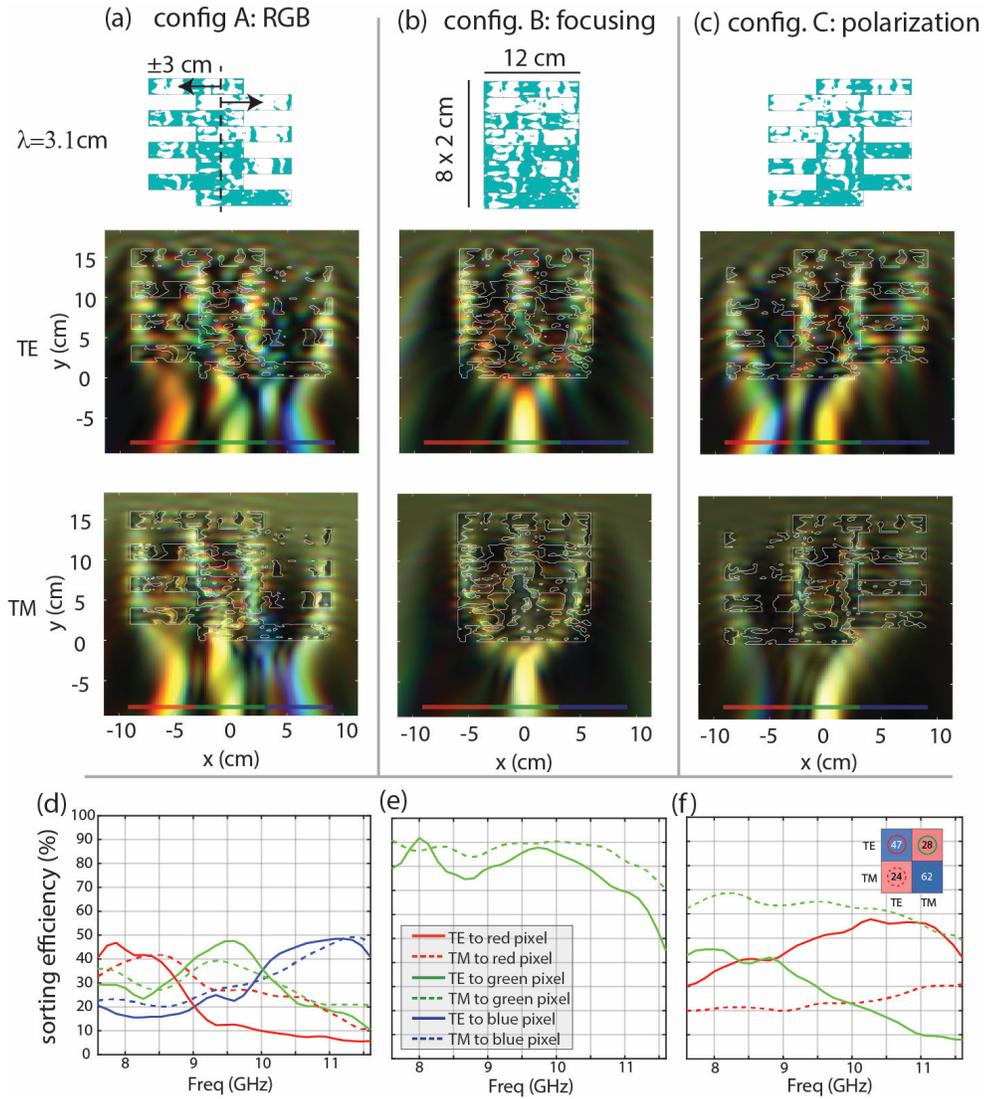

Fig. 4. TE and TM fields for a device based on a net shearing movement of 6 cm. (a) The spectral splitting configuration splits red, green, and blue light to the red, green, and blue pixels, respectively. This is analyzed for both TE and TM polarizations. (b) The neutral state of the device performs broadband focusing to the center pixel for both TE and TM polarizations. (c) The polarization sorting configuration sorts broadband TE light to the left (red) pixel and focuses broadband TM light to the center (green) pixel. (d-f) Sorting efficiency of the different configurations. The line color represents the sorting efficiency to the similarly colored pixel as depicted in the color plots. Solid lines represent the TE response, and dashed line represent the TM response. (d) Sorting efficiency in the spectral splitting configuration. (e) Sorting efficiency for the broadband focusing configuration. (f) Sorting efficiency for the polarization splitting configuration. (Inset) A confusion matrix representation of the sorting efficiency, with true input on the vertical axis and predicted input on the horizontal axis. Each matrix entry is determined by averaging the relevant trace over the full bandwidth.

## 4. Discussion

We have studied dielectric scattering elements capable of substantially expanding their functionality through mechanical reconfiguration. The work here is a step towards answering an important question in optics – how much performance and functionality can be squeezed into a certain volume? The miniaturization of optical systems has led to new applications in lasers [25,26], biomedical optics [27,28], space instrumentation [29], and generally applications where strict size requirements exist. Yet there is still much progress that can be made, since systems of cascaded metasurfaces still generally require some free-space propagation to preserve their independence and the validity of the techniques often employed in metasurface design. Furthermore, complex functionality such as combining polarization control, spectral splitting, and imaging into single elements has been elusive.

The approach shown here showcases the advantages of coupling mechanical reconfiguration with adjoint-based electromagnetic optimization. Miniaturization is achieved by employing adjoint-based inverse design to find the optimal shape of a small dielectric volume, while broad multifunctional performance is achieved through mechanical reconfiguration. The designed structures are difficult to fabricate at the nanoscale for optical applications, but the design process is flexible enough to incorporate nanoscale fabrication requirements as shown in other works [30]. In particular, some areas of fabrication that can fabricate the required subwavelength features for these devices include: multilayer fabrication common in CMOS or MEMS processes [31]; direct-write two-photon laser lithography capable of designing subwavelength 3D elements in the infrared [17]; and closely aligned stacks of Silicon wafers with subwavelength features at terahertz frequencies [32]. All of these techniques have demonstrated active mechanical control that could be useful for multiplexing functions like what was demonstrated in this work.

Adjoint-optimization is well-suited for optimizing the optical properties of mechanically reconfigurable devices. In this work, the mechanical reconfiguration scheme was predetermined, and the material was patterned to enable multifunctionality. However, future work could combine existing optimization techniques for designing mechanical metamaterials [33,34] with the techniques presented here, yielding devices that have co-optimized mechanical and optical performance.

Acknowledgements: This work was funded by Defense Advanced Research Projects Agency EXTREME program (HR00111720035).